\documentclass[twocolumn,tighten]{aastex61} 
\usepackage{graphicx}
\usepackage{amsmath}
\usepackage{natbib}
\usepackage{pdfpages}
\usepackage[textheight=540pt]{geometry}

\submitjournal{ApJ}
\shortauthors{Harris \& Mulholland}

\begin{document}\sloppy

\title{Detection of the Stellar Intracluster Medium in Perseus (Abell 426)}

\correspondingauthor{William E.~Harris}
\email{harris@physics.mcmaster.ca}

\author{William E.~Harris}
\affiliation{Department of Physics \& Astronomy \\
McMaster University \\
Hamilton ON L8S 4M1, Canada}

\author{Courtney J.~Mulholland}
\affiliation{Department of Physics \& Astronomy \\
McMaster University \\
Hamilton ON L8S 4M1, Canada}

\date{\today}

\begin{abstract}
Hubble Space Telescope photometry from the ACS/WFC and WFPC2 cameras is used
to detect and measure globular clusters (GCs) in the central region of 
the rich Perseus cluster of galaxies.  A detectable population of Intragalactic GCs is
found extending out to at least 500 kpc from the cluster center.  These objects display
luminosity and color (metallicity) distributions that are entirely normal for GC populations.
Extrapolating from the limited spatial coverage of the HST fields,
we estimate very roughly that the entire Perseus cluster should contain $\sim 50000$ or more IGCs, 
but a targetted wide-field survey will be needed
for a more definitive answer.  Separate brief results are presented for the rich GC systems in
NGC 1272 and NGC 1275, the two largest Perseus ellipticals.  For NGC 1272 we find
a specific frequency $S_N \simeq 8$, while for the central giant NGC 1275, $S_N \simeq 12$. 
In both these giant galaxies, the GC colors are well matched by bimodal distributions, with the
majority in the blue (metal-poor) component.  This preliminary study suggests that Perseus
is a prime target for a more comprehensive deep imaging survey of Intragalactic GCs. 
\end{abstract}


\section{Introduction}

In rich clusters of galaxies, Intra-Cluster Light (ICL) is often present, consisting of stars
that are associated with the extended potential well of the cluster as a whole, rather than
the individual galaxies.  The ICL can be produced through the merger, tidal stripping, and
harassment events that happen whenever close encounters between the member galaxies take place,
and in rich clusters the ICL component can be substantial in total even if sparsely spread
\citep[e.g.][]{purcell_etal2007,martel_etal2012,contini_etal2014,cooper_etal2015}.

Much observational work on Abell-type clusters and compact groups through low-surface-brightness photometry
confirms the frequent presence of an ICL usually well centered on the BCG (Brightest Cluster
Galaxy)
\citep[e.g.][among many]{gonzalez_etal2005,darocha_mendes2005,presotto_etal2014,bender_etal2015,iodice_etal2016}; see \citet{peng_etal2011} for a more complete listing of
earlier literature.  Consistent with its probable origin by stripped stars from all types of
cluster galaxies, population synthesis shows that the ICL contains large numbers of old, relatively
metal-rich stars such as are in giant ellipticals, but 
also a significant presence of younger, more metal-poor stars from smaller galaxies
\citep{coccato_etal2011,melnick_etal2012,demaio_etal2015,edwards_etal2016,barbosa_etal2016}.  
Assessing the true fraction of the
ICL component relative to the individual galaxies is a continuing and difficult issue since it
requires careful separation of the ICL profile from the giant galaxies in the cluster,
but is almost certainly in the range 10-50\% for rich clusters and even groups
\citep{gonzalez_etal2005,darocha_mendes2005,seigar_etal2007,rudick_etal2011,giallongo_etal2014,jimenez-teja_dupke2016}.
The growth of the ICL is an ongoing process with much of it happening since redshift
$z \sim 1$ \citep[][]{burke_etal2012,burke_etal2015,adami_etal2013,presotto_etal2014}, \
so the total luminosity in the ICL and its degree
of substructure (clumps and tidal streams) are indicative of the dynamical evolutionary
state of the cluster.

For nearby cluster environments (Virgo, Fornax, Coma) the ICL has been probed through resolution
into individual stars including red-giant stars \citep{ferguson_etal1998,williams_etal2007}, 
planetary nebula 
\citep{feldmeier_etal2003,arnaboldi_etal2007,castro-rodrigues_etal2009,ventimiglia_etal2011,longobardi_etal2013}, and even
novae and supernovae \citep{neill_etal2005,gal-yam_etal2003,mcgee_balogh2010,sand_etal2011,graham_etal2015}.
But another prominent tracer of the ICL that has so far been underutilized is the globular
cluster (GC) population, which accompanies all types of old stellar populations in galaxies.
GCs are individually bright and can readily be measured in and around galaxies that are as
distant as $d \sim 300$ Mpc \citep[e.g.][]{harris_etal2014} especially with HST imaging.
\emph{Intragalactic globular clusters} (IGCs) in large numbers have clearly been found 
populating the Virgo cluster by means of wide-field ground-based surveys 
\citep{lee_etal2010,durrell_etal2014} and four individual Virgo IGCs have been studied 
through deep HST imaging \citep{williams_etal2007}.  More tentative detections of IGCs have been
made for the Fornax cluster \citep{bassino_etal2003,bergond_etal2007,schuberth_etal2008}
and a targetted wide-field survey may reveal its IGC distribution more clearly 
\citep{dabrusco_etal2016}.  In the richer Coma cluster, \citet{peng_etal2011} find a
substantial IGC population filling the cluster core regions.
Three other and more distant galaxy clusters for which IGCs have been sampled
with unusually deep HST imaging include A1185 \citep{west_etal2011}, A1689 \citep{alamo-martinez_etal2013}, and
A2744 \citep{lee_jang2016}, though for various reasons of either field coverage or depth these
also remain more uncertain than for Virgo or Coma \citep[see also][for additional discussion of A1689 and 
A2744]{harris_etal2017}.

The Perseus cluster (Abell 426) is another very rich system whose GC population is well within reach of the HST
cameras or even deep ground-based imaging under good conditions.  At a distance $d = 75$ Mpc 
(taking a mean redshift $cz = 5207$ km s$^{-1}$ (NED) and
an adopted distance scale $H_0 = 70$ km s$^{-1}$ Mpc$^{-1}$), it is 0.5 mag closer than the Coma cluster 
and contains a comparably rich set of member galaxies including many large ellipticals.  
Perseus is embedded in a vast halo of X-ray gas filling the cluster
\citep{urban_etal2014} which has a virial radius $r_{200} =$ 1.8 Mpc. 
However, virtually nothing is known about any \emph{stellar} component
of the Perseus Intragalactic Medium (IGM).

NGC 1272 ($M_V^T = -22.96$) and NGC 1275 ($M_V^T = -22.61$) are the two dominant member galaxies of Perseus
and have similar visual luminosities. In this respect Perseus is analogous to several other rich clusters
where two major E galaxies of roughly equal luminosity dominate (for example, NGC 4472 and 4486 in Virgo, or
NGC 4874 and 4889 in Coma).  However,
NGC 1275 appears to be the closest to the dynamical center of Perseus, based on its
projected location virtually at the center of the hot X-ray halo,
its radial velocity (5264 km s$^{-1}$) close to the cluster mean,
and the dramatic presence of large gas inflow revealed in its
extensive gas and dust network of filaments.  Although NGC 1272 is also projected fairly close
to cluster center, its radial velocity (3815 km s$^{-1}$) is much further from the cluster mean,
and its integrated-light profile is smooth and featureless at all radii. 

In this paper, we take advantage of Archival HST images near the center of Perseus to search
specifically for IGCs as a tracer for the ICL.  Our finding is that a substantial IGC
population is present, separate from those associated with any of the major galaxies.

For $d = 75$ Mpc the distance modulus of Perseus is $(m-M)_0 = 34.36$.  For the following
analysis we adopt a foreground
absorption $A_V = 0.447, A_I = 0.245$ from NED, which gives $E_{V-I} = 0.202$.

\section{Data and Photometric Measurement}

\begin{figure}[t]
\vspace{-0.0cm}
\begin{center}
\includegraphics[width=0.4\textwidth]{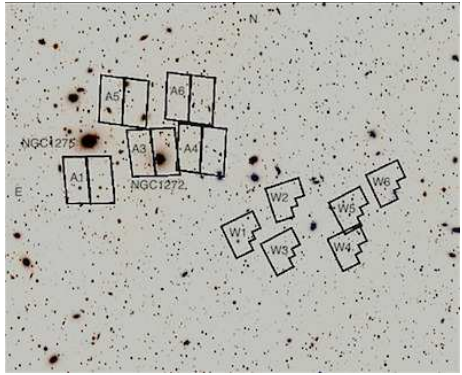}
\end{center}
\vspace{-0.2cm}
\caption{Image of the central region of the Perseus cluster (DSS image in blue IIIaJ), 
	showing the locations of the HST images used in this study.
North is at top, East at left.  
The field shown is $36'$ across, corresponding to 750 kpc at the distance of Perseus.
Note that field ACS-1 is below the central cD NGC 1275, while field ACS-3 contains NGC 1272.}
\vspace{0.0cm}
\label{fig:fields}
\end{figure}

The data for this study are drawn from from two HST imaging programs:  5 ACS/WFC fields from
program GO-10201, and 6 WFPC2 fields from program GO-10789 (PI Conselice).  In both cameras
the images were taken in the $F555W$ and $F814W$ bands, thus putting the photometry close to standard $(V,I)$.
Previous photometric analysis of these images is directed primarily at detecting and
assessing the UCDs (Ultra-Compact Dwarfs) in Perseus \citep{penny_etal2011,penny_etal2012}.
The locations of these ACS and WFPC2 fields are shown in Figure \ref{fig:fields}
and the locations are listed in Table \ref{tab:coords}
\citep[see also Fig.~1 of][]{penny_etal2011}.  The ACS fields cover parts of the Perseus core
with several large galaxies nearby;
ACS-1 is located just below the central cD NGC 1275, while the giant elliptical NGC 1272
falls within ACS-3. The WFPC2 fields scatter towards 
the west, and are farther from any of the large Perseus member galaxies.

\begin{table*}[t]
\begin{center}
\caption{\sc Target Fields}
\label{tab:coords}
\begin{tabular}{lllcccc}
\tableline\tableline\\
\multicolumn{1}{l}{Field} &
\multicolumn{1}{l}{RA} &
\multicolumn{1}{l}{Dec}  &
\multicolumn{1}{c}{$\alpha(F814W)$} &
\multicolumn{1}{c}{$m_0(F814W)$} &
\multicolumn{1}{c}{$\alpha(F555W)$} & 
\multicolumn{1}{c}{$m_0(F555W)$} 
\\[2mm] \tableline\\
ACS-1 & 03 19 47.91 & +41 28 00.27 & 3.75 & 27.39 & 3.75 & 28.10 \\
ACS-3 & 03 19 23.18 & +41 29 59.10 & 3.50 & 27.47 & 3.50 & 28.10 \\
ACS-4 & 03 19 04.73 & +41 30 10.93 & 4.00 & 27.43 & 4.00 & 28.17 \\
ACS-5 & 03 19 34.58 & +41 33 37.23 & 3.75 & 27.39 & 4.00 & 28.02 \\
ACS-6 & 03 19 09.14 & +41 33 50.43 & 3.75 & 27.45 & 3.75 & 28.10 \\
\\
WFPC2-1 & 03 18 48.75 & +41 24 06.75 & 3.00 & 25.41 & 2.25 & 26.28 \\
WFPC2-2 & 03 18 32.26 & +41 26 36.13 & " & " & " & " \\
WFPC2-3 & 03 18 33.55 & +41 22 53.43 & " & " & " & " \\
WFPC2-4 & 03 18 08.05 & +41 23 10.83 & " & " & " & " \\
WFPC2-5 & 03 17 53.35 & +41 27 45.68 & " & " & " & " \\
WFPC2-6 & 03 17 30.15 & +41 25 09.89 & " & " & " & " \\
\\[2mm] \tableline
\end{tabular}
\end{center}
\vspace{0.4cm}
\end{table*}

\begin{figure}[t]
\vspace{-0.0cm}
\begin{center}
\includegraphics[width=0.4\textwidth]{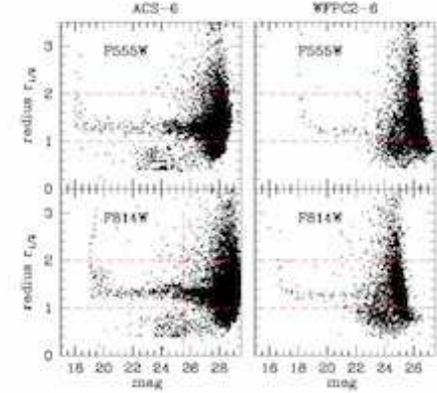}
\end{center}
\vspace{-0.2cm}
\caption{\emph{Left panels:} SourceExtractor parameters $r_{1/2}$ versus aperture magnitude,
	for both filters in ACS field 6.  \emph{Right panel:}  The same quantities for
	WFPC2 field 6.  Units of $r_{1/2}$ are pixels, where 1 px = $0.05''$ for ACS
	and $0.1''$ for WFPC2. Objects in the `starlike' range $1 < r_{1/2} < 2$ are selected for
further photometry with \emph{daophot/allstar}.  Vertical dashed lines show the $F814W$
magnitude limit adopted for analysis of the GC populations (see text).}
\vspace{0.0cm}
\label{fig:se}
\end{figure}

\begin{figure*}[t]
\vspace{-0.0cm}
\begin{center}
\includegraphics[width=0.7\textwidth]{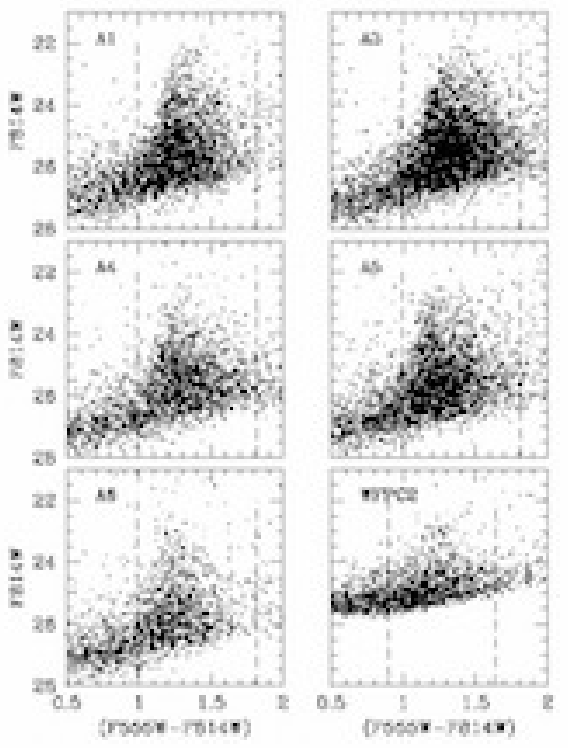}
\end{center}
\vspace{-0.2cm}
\caption{Color-magnitude diagrams for the measured unresolved (starlike) objects in the Perseus fields.
	The five ACS/WFC fields are labelled A1,3,4,5,6, while the six WFPC2 fields are combined
	into one CMD at lower right.  The vertical dashed lines at $(F555W-F814W) = 0.99, 1.81$
	(for ACS) and $0.89, 1.64$ (for WFPC2) mark the (generous) range of colors populated by normal globular
	clusters. The detection limit ($m_0$ in Table 1) for the WFPC2 fields is clearly much brighter than for the ACS fields
because of shorter exposure times and lower detector sensitivity.}
\vspace{0.0cm}
\label{fig:cmd6}
\end{figure*}

At 75 Mpc distance, a typical GC half-light diameter $\simeq 4$ pc converts to 
an image FWHM of 0.01 arcsec, an order of magnitude smaller than the resolution limit of
the HST cameras \citep[see also][for discussion]{harris2009a}.  
The great majority of the GCs in the Perseus galaxies will
therefore be comfortably starlike in appearance, which greatly simplifies the process of
GC candidate object selection and photometry.  Essentially, we look for \emph{unresolved}
objects within each ACS or WFPC2 field that also fall within the (fortunately)
relatively narrow color range in $(V-I)$ that match real GCs.  As will be seen
below, the combination of image morphology and color eliminates almost all contaminants
in the sample (foreground field stars, or very faint small background galaxies).

Data reduction began with the \emph{*.flc} image files from the HST Archive.  For each of the
target fields, the individual exposures (4 in each filter) were registered and combined with
\emph{stsdas/multidrizzle} into a single composite exposure in each filter.  
Total exposure times for each ACS/WFC field are 2260 s ($F814W$) and 2368 s ($F555W$).
For each of the 6 WFC2 fields, total integration times in each filter
were 1900 s.

The procedure for detection and photometry of the GC candidates then followed the steps
outlined in, e.g., \citet{harris2009a,harris_etal2016}.
SourceExtractor (SE) \citep{bertin_arnouts1996} was used to detect objects
and to do a preliminary rejection of nonstellar ones (either single-pixel artifacts
or faint background galaxies) from a plot of the SE parameters $r_{1/2}$ versus aperture magnitude.
Samples of these for one ACS field and one WFPC2 field are shown in Figure \ref{fig:se}.
Starlike (point-source) objects fall near $r_{1/2} \simeq 1.3$ px and objects 
with $r_{1/2} < 1$  or $> 2$ in \emph{either} filter were rejected.
Final photometry of the candidates was done through the tools in \emph{iraf/daophot},
including point-spread-function (PSF) fitting and \emph{allstar} solutions, then corrected
to equivalent large-aperture magnitudes.
The PSF in each field was determined empirically from moderately bright, uncrowded stars.
The object lists in each filter were matched to within $0.1''$ and any objects not appearing
in both filters were rejected.  Finally, objects with poor goodness-of-fit to the PSF 
($\chi > 2$) or large photometric uncertainties ($\sigma > 0.3$ mag) were also rejected.
The combination of steps (SE parameters, $\chi$ and $\sigma$ rejection, matching lists
in both filters) proved to be quite effective at culling nonstellar objects.
We note that rejection of nonstellar objects is not expected to be as rigorous 
on the WFPC2 fields, which have a larger pixel size than ACS/WFC ($0.1''$ versus $0.05''$)
and shorter exposure times as well.  
Within the WFPC2 array we used the WF2, WF3, and WF4 detectors but not the small PC1,
to avoid issues with different limiting magnitudes and resolution.

Photometric completeness as a function of magnitude was evaluated through \emph{daophot/addstar}
tools, where scaled PSFs are added to the images and then the process of detection and
photometry is repeated.  Since every one of the target fields is quite uncrowded in 
absolute terms, artificial-object lists were generated randomly in location across the
field.  The dependence of completeness $f(m)$ (the fraction of successfully detected and
measured objects versus magnitude) is modelled with the simple two-parameter function
\citep{harris_etal2016} $f = (1 + \mathrm{exp}(\alpha(m-m_0))^{-1}$, where $m_0$ is the magnitude
at which 50\% of the objects are detected and $\alpha$ measures the steepness of falloff
of the completeness curve.  In Table \ref{tab:coords} we list the $(\alpha,m_0)$ pairs
for each ACS field individually; as expected, they are closely similar from one field
to the next.  For the WFPC2 fields we list the average of the 6 fields.

In the following discussion, we present the
color-magnitude data in the natural filter magnitudes $(F555W,F814W)$. 
For ACS/WFC these can be transformed to $(V,I)$ through \citep{saha_etal2011}
\begin{eqnarray}
	(V-I) \, = \, 0.912(F555W-F814W) \, \\
	I \, = \, F814W + 0.024 (V-I)
\end{eqnarray}
while for WFPC2 the adopted transformation is \citep{holtzman_etal1995} 
\begin{eqnarray}
	(V-I) \, = \, 1.010(F555W-F814W) \, \\
	I \, = \, F814W - 0.062 (V-I) \, .
\end{eqnarray}

\section{Analysis}

\subsection{Color-Magnitude Distributions}

In Figure \ref{fig:cmd6} the distribution of measured unresolved (starlike) objects in
the color-magnitude diagram (CMD) is shown, for each of the five ACS fields and for
all six WFPC2 fields combined.  In the $(I,V-I)$ plane, normal GCs will fall well within a 
generous color range $(V-I)_0 \simeq 0.70 - 1.45$ \citep{larsen_etal2001} where
color is primarily a function of cluster metallicity.  Thus for
our adopted reddening $E_{V-I} = 0.202$, this range of intrinsic colors translates into
$(F555W-F814W) \simeq 0.99-1.81$ (ACS) or $\simeq 0.89-1.64$ (WFPC2).

In every ACS field a substantial population appears in this target range, dominating all
objects with $I \lesssim 26$. (Note that the swath of points along the bottom running 
diagonally up from lower left
is at or below the completeness limit of the photometry and thus 
mostly represent extremely uncertain detections or noise).  Adding to the evidence for
their identification as GCs is that the bright end effectively stops near $I \simeq 22$, which
translates to $M_I = -12.5$ or roughly $M_V = -11.5$ for a mean intrinsic color
$(V-I)_0 = 1.0$.  The luminosity distribution of normal GCs (GCLF) is highly
consistent between galaxies, and reaches a ``top end'' at $M_I \simeq -12.5$
even for the most luminous galaxies with the biggest GC populations
\citep{harris_etal2014}.

Looking further into the CMDs, in most galaxies the GC population falls into a bimodal distribution where ``blue''
(metal-poor) ones peak near $(V-I)_0 \simeq 0.95$ and ``red'' (metal-richer) ones
near $(V-I)_0 \simeq 1.15$ \citep[e.g.][]{larsen_etal2001}.  For the Perseus fields
these translate to mean colors $(F555W-F814W) = 1.25$ and 1.50 (ACS filters).  
Both the blue and red sequences can be seen at these colors in the five ACS fields.
The actual color distribution functions will be discussed in more detail below.

In the combined WFPC2 fields (lower right panel of Fig.~\ref{fig:cmd6}), the scatter
	of the photometry is larger and the limiting magnitudes shallower.
	Nevertheless, a thinly populated sequence of objects with $F814W < 24$ 
	and near $(F555W-F814W) \simeq 1.2$
	is clearly present.  This mean color is equivalent to $(V-I)_0 = 1.0$, exactly
	the mean intrinsic color of normal GCs.  The presence of this sequence
	is the first evidence that these outlying fields have a sparsely spread
GC population.

\subsection{Luminosity Distributions}

\begin{figure}[t]
\vspace{-0.0cm}
\begin{center}
\includegraphics[width=0.4\textwidth]{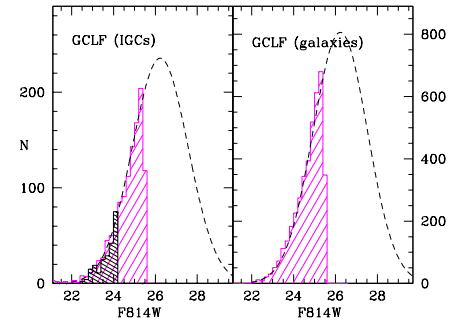}
\end{center}
\vspace{-0.2cm}
\caption{Number of objects per 0.2-mag bin in $F814W$, plotted versus
	magnitude.  \emph{Left panel:} The \emph{magenta} histogram gives the numbers
		for the fields ACS-4 and ACS-6 combined, which should be dominated
		by IGCs. For $F814W < 25.5$ the photometric completeness is near 100\%.
		The black shaded histogram gives the numbers for the 6 outlying WFPC2 fields combined,
		over the range $F814W < 24.2$ where photometric completeness is high.
		The Gaussian curve shown as the dashed line has a turnover
	(peak) at $F814W = 26.2$ and standard deviation 1.33 mag.
\emph{Right panel;} The magenta histogram gives the numbers per 0.2-mag bin
in the fields ACS-1,3,5 combined; these are dominated by the giant ellipticals
in the Perseus core.  The Gaussian dashed line has a standard deviation 1.30 mag
and the same turnover as in the left panel.}
\vspace{0.0cm}
\label{fig:lf}
\end{figure}

A further test of `normal' old GC populations is that they should follow a Gaussian-like
distribution in number per unit magnitude \citep[e.g.][]{jordan2007,villegas2010,harris_etal2014}.
For the Perseus data, we anticipate that fields ACS-1,3,5 should be predominantly 
populated by GCs from the giant ellipticals in the Perseus core,
while ACS-4,6 and the WFPC2 fields should have a much higher fraction of IGCs if
they are present (see next section for the spatial distributions).  In Figure \ref{fig:lf},
we show the luminosity functions for these three groups of fields separately.

A typical GCLF based on the data for individual galaxies of all luminosities
(see the references cited above) should have a Gaussian turnover (peak) luminosity
$M_V \simeq -7.3$, $M_I \simeq -8.3$ (assuming a mean $(V-I) \simeq 1.0$),
and a standard deviation $\sigma \simeq 1.3$ mag.  
Both the turnover and width of the fiducial GCLF are empirically
uncertain by $\sim 0.1$ mag \citep[e.g.][]{jordan2007}. 
For Perseus, the GCLF turnover would then be expected to lie
at $I \simeq 26.2$, which is fainter than the useful photometric limits
in the ACS fields and considerably fainter than the WFPC2 field limit.  

Since the data fall short of the turnover magnitude $m_0$, both $m_0$ and $\sigma$ cannot
be solved simultaneously very accurately \citep[e.g.][]{hanes_whittaker1987}, so instead we
simply assume $m_0(F814W) = 26.2$ and solve for $\sigma$ as a consistency test of the
Gaussian form.  To do this, we used the objects in a restricted  
magnitude range $F814W < 25.5$ (for the ACS fields) within
which the photometric completeness is nearly 100\% and
the CMD is minimally affected by either field contamination or spread of
photometric  errors.  The histograms of number per 0.2-mag bin in that range are
shown in Fig.~\ref{fig:lf}.  Least-squares fits of a Gaussian curve yield
$\sigma = 1.30 \pm 0.03$ mag for ACS-1+3+5 (the fields dominated by Perseus core galaxies),
and $\sigma = 1.33 \pm 0.04$ mag for ACS-4+6.  In both cases the model curve fits the data
closely.  In the left panel of Fig.~\ref{fig:lf} the
LF for the combined WFPC2 fields is also shown, indicating that these outer regions are
also consistent with the assumption of a normal GCLF.

\subsection{Background}

Assessing whether or not we are looking at an IGC population, particularly for the 
sparse outer WFPC2 fields,
also requires having some idea of the level of field contamination.
The best way to do this would be through control fields from WFC3, ACS/WFC, or WFPC2 located close
to but outside Perseus and with the same filters and similar exposure times. 
Unfortunately a search of the HST Archive did not yield any useful cases.

\begin{figure}[t]
\vspace{-0.6cm}
\begin{center}
\includegraphics[width=0.4\textwidth]{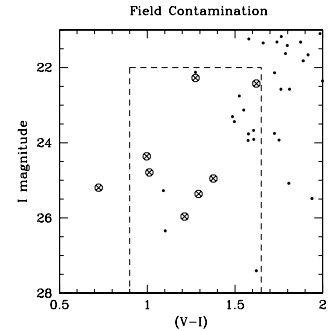}
\end{center}
\vspace{-0.6cm}
\caption{\emph{Solid dots:} Color-magnitude diagrams for a simulated Milky Way population of stars in the
	direction of NGC 1275, generated by TRILEGAL.  This example is for an area
on the sky equal to 5 ACS/WFC fields combined.  The dashed lines enclose the
expected range of magnitude and colors for GCs in the Perseus galaxies.
\emph{Circled crosses:}  Starlike objects in the HUDF (see text).}
\vspace{0.0cm}
\label{fig:trilegal}
\end{figure}

One alternative is to estimate the expected number of foreground stars with a Milky Way 
stellar population model.  In Figure \ref{fig:trilegal}, we show a sample CMD produced
by TRILEGAL \citep{girardi_etal2005} in the direction of NGC 1275 and comprising an area
equal to all 5 ACS fields combined.  Since the projected area on the sky is small, 
different random realizations of TRILEGAL show large stochastic differences, but the
example shown here is typical.  Only $\sim 10$ stars appear within the broad magnitude
and color range for the Perseus GCs, and most of these are too bright and red to fall
within the normal GC sequences described above.  

Contamination may also arise from faint, very compact background galaxies that 
slip through the initial rejection criteria (SE parameters, $\chi$, magnitude
uncertainty, and color range).  To estimate the numbers of such objects
that we would expect per ACS field, data from the Hubble Ultra-Deep Field (HUDF) 
were used as provided from the catalogs at
heasarc.gsfc.nasa.gov/W3Browse/hst/hubbleudf.html.
Fig.~\ref{fig:trilegal} shows the HUDF objects that can be classified as starlike
($r_{1/2} < 2$ px from SourceExtractor) and fall in the color range 
of interest.  (Note that these objects were made fainter and redder by
$A_V = 0.447, E_{V-I} = 0.202$ to match the Perseus field.)  
The vast majority of `starlike' objects in the HUDF are much fainter than
our target range, and most of the rest are too blue or red to match GC colors.
The remainder consist of only a handful of contaminants with $I \gtrsim 24.5$
and within the color range marked in Fig.~\ref{fig:trilegal}.
Much the same result was found in a very similar study of the GC population
around NGC 6166 \citep{harris_etal2016} (see especially Figs.~8 and 9 there).

For each of the ACS fields, the field contamination should then
make up less than 1\% of the observed numbers of objects
in our CMDs (Fig.~\ref{fig:cmd6}), so we do not remove any background
from the LFs or the radial distributions described below.
For the WFPC2 fields, as seen for the discussions of the luminosity
function and color distribution function, we use only the objects with $I < 24.0$,
which further excludes the contaminating objects in Fig.~\ref{fig:trilegal}
almost totally.

In summary, the current material points to the presence of a significant population of \emph{bona fide}
GCs around the Perseus galaxies.   These
objects follow the distributions expected for normal GCs in terms of their
magnitude, color, and unresolved size.  We can now continue to a discussion of their
	spatial distribution within Perseus and relative to the two dominant galaxies,
NGC 1272 and 1275. 

\subsection{Spatial Distributions}

As seen in Fig.~\ref{fig:cmd6}, the largest GC populations are in fields ACS-1 (which is nearest NGC 1275)
and ACS-3 (which includes NGC 1272).  The star-cluster population in NGC 1275 itself has been
discussed briefly by \citet{carlson_etal1998}, \citet{canning_etal2010}, and 
\cite{penny_etal2012}, derived from WFPC2 and ACS images centered on the galaxy 
(though in different filters) that we do not
use here.  In total, the NGC 1275 star clusters consist of a
mixture of old GCs spread throughout the galaxy's halo, plus a rich population of younger
clusters located along its spectacular network of H$\alpha$ filaments.  

\begin{figure*}[t]
\vspace{-0.0cm}
\begin{center}
\includegraphics[width=0.6\textwidth]{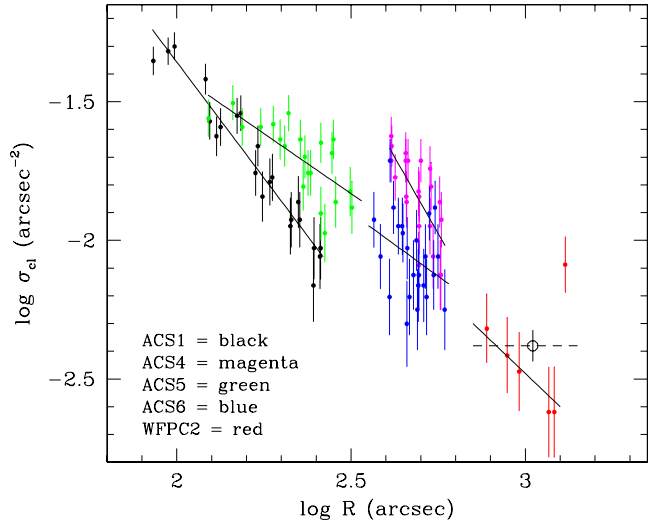}
\end{center}
\vspace{-0.2cm}
\caption{Number of objects per arcsec$^2$, plotted versus radial
	distance from NGC 1275 (the Perseus center).  To be used in this diagram,
	a measured object needed to be (a) starlike, (b) in the color range shown
	in Fig.~\ref{fig:cmd6}, and (c) brighter than $F814W = 25.5$ (see text).  The data are
	color-coded by field location.  For each ACS field the best-fitting power law curve
	$\sigma_{cl} \sim R^{\alpha}$ is drawn in.  For the WFPC2 fields (in red symbols),
	the mean point (large open symbol) is the total number in all 6 fields divided by
their total area.}
\vspace{0.0cm}
\label{fig:radial}
\end{figure*}

Since NGC 1275 is the central giant, we 
first plot the GC number density versus radial distance from it.
The results are shown in Figure \ref{fig:radial}.  Here, the number of GCs 
per unit projected area on the sky $\sigma_{cl}$ 
is calculated within radial annuli centered on NGC 1275.  For the ACS fields, the GCs used in the sums
include objects with $F814W < 25.5$, $1.0 < (F555W-F814W) < 1.8$.  
At the adopted limit $F814W(lim) = 25.5$, the internal photometric errors in the
ACS data are $e_I = \pm 0.10$, $e_{V-I} = \pm 0.13$. 

Placing the raw WFPC2 counts on the same
graph as the ACS fields requires scaling them to the same magnitude limit.  
To do this, we truncate the adopted limiting magnitude now more severely to $F814W < 23.5$,
$0.9 < (F555W-F814W) < 1.65$ to avoid the fainter
cloud of points which may be dominated
by small faint background galaxies and even artefacts that escaped the photometric culling.  
At $F814W = 23.5$ the internal photometric errors
are $e_I = \pm 0.06$, $e_{V-I} = \pm 0.11$.  A useful test of contamination
is the field-to-field scatter in the numbers of detected
objects.  For $F814W(lim) = 24.2$, we find that the scatter amongst the 6 WFPC2 fields
is much larger than the $\sqrt{N}$ level expected
from simple count statistics, which would be the expected result from the known physical clustering of faint background
galaxies over several arcminutes scale.  By comparison, for $F814W(lim) \lesssim 23.5$, the field-to-field scatter
is near $\sqrt{N}$, indicating that most of the contaminating galaxies are 
gone.\footnote{The actual numbers of objects brighter than $F814W=23.5$ and
within the stated color range in the WFPC2 fields $W1 - W6$ are 10, 8, 7, 5, 5, 17.
We note that reducing the adopted cutoff magnitude even further to (say) $F814W(lim) = 23.0$ leaves 
extremely small-number statistics and large normalization factors 
from which it is difficult to draw any conclusions.} 
From the ACS counts, there are 1210 objects with $F814W \leq 25.5$
and within the desired color range, but only 131 with $F814W \leq 23.5$.  The WFPC2 counts
are therefore scaled up by the ratio ($9.24 \pm 0.85$) to give the equivalent number brighter than
$F814W=25.5$.  These renormalized $\sigma-$values for the individual WFPC2 fields 
are plotted in Fig.~\ref{fig:radial} along with the value from all 6 fields combined.

In Fig.~\ref{fig:radial} the different regions are color-coded for easier distinction:

\emph{ACS-1} shows the outer NGC 1275 halo, and the strong radial falloff of $\sigma_{cl}$ 
relative to its center is evident.  The power-law slope shown in the Figure is well defined
by the data and has slope $\alpha = (\Delta {\rm log} \sigma_{cl} / \Delta {\rm log} R) = -1.67 \pm 0.01$.
This deduced slope agrees well with $\alpha = -1.6$ found by 
\citet{penny_etal2012} based on both ACS-1 and ACS/WFC images centered on the galaxy.

\emph{ACS-3} encloses the giant NGC 1272 and
its GC population is clearly dominated by it \citep[see][]{canning_etal2010,penny_etal2012}, so it
is not plotted in Fig.~\ref{fig:radial} relative to the center of NGC 1275.
It will be discussed in Section 4 below.

\emph{ACS-4} may be a combination of NGC 1272 halo and IGC; for it, we find
a radial falloff $\alpha = -2.20 \pm 0.50$.

\emph{ACS-5}, from its location and $\sigma_{cl}$ behavior, suggests that we are
looking at a combination of populations from the outer halos of both NGC 1275 and NGC 1272,
plus some less certain contributions from the smaller ellipticals in the field and the IGCs.
The formal solution for the radial falloff gives $\alpha = -0.87 \pm 0.19$.

\emph{ACS-6} is farthest from both
the major galaxies, so the IGC component should be relatively more dominant.  A shallow
and uncertain radial falloff is seen with $\alpha = -0.9 \pm 0.4$.

\emph{WFPC2:} the WFPC2 fields contain no major Perseus galaxies 
and are located radially from $R=270$ to 530 kpc
from the Perseus center, and should therefore give us a fairly clean IGC sample (though
unfortunately to shallower magnitude limits, as discussed above).  Interestingly, the 6 points 
in Fig.~\ref{fig:radial} show an internally uncertain but distinctive downward radial gradient, with
the exception of the outermost W6.  The number density for all 6 of these fields
combined is $\langle \sigma_{cl} \rangle = (0.0042 \pm 0.0007)$ arcsec$^{-2}$, shown
by the large open symbol and errorbar.  The solid line drawn through the points
has a slope $\alpha \simeq -1.2$ and seems at least roughly to continue the outward trend
from ACS-6.  Though the field coverage is admittedly only a small fraction of the
Perseus core and outskirts, these data hint that we may be observing the radial
distribution of the IGC component, far from any major member galaxies.

\begin{figure}[t]
\vspace{-0.5cm}
\begin{center}
\includegraphics[width=0.5\textwidth]{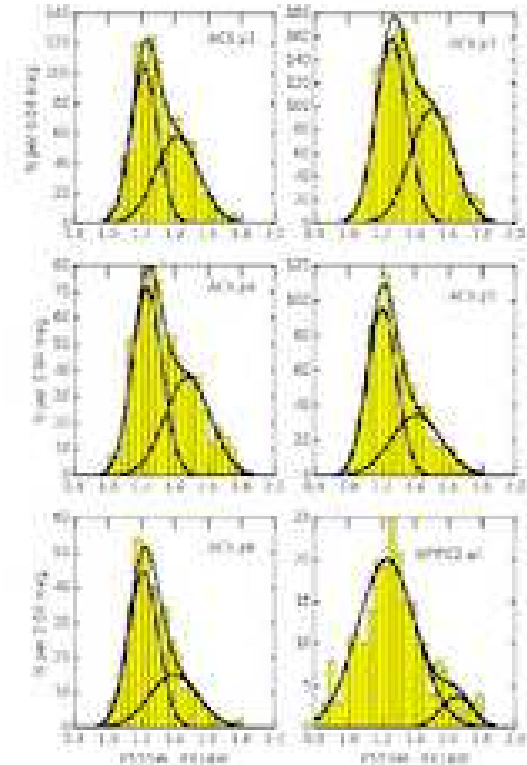}
\end{center}
\vspace{-0.0cm}
\caption{GMM fits to the color distribution functions in the five ACS fields and
(at lower right) the combined data from the WFPC2 fields.
The histograms in yellow shows the number of objects in bins of width 0.04 mag
in color.  In each panel, the solid lines show the best-fit blue and red modes, and the sum 
of the two modes as the upper envelope.}
\vspace{0.0cm}
\label{fig:gmm}
\end{figure}

\subsection{Metallicity and Color Distributions}

\begin{table*}[t]
\begin{center}
\caption{\sc Bimodal Gaussian Fits to the CDF}
\label{tab:gmm}
\begin{tabular}{lcccccc}
\tableline\tableline\\
\multicolumn{1}{l}{Field} &
\multicolumn{1}{c}{$\mu_1$} &
\multicolumn{1}{c}{$\mu_2$} &
\multicolumn{1}{c}{$\sigma_1$}  &
\multicolumn{1}{c}{$\sigma_2$} &
\multicolumn{1}{c}{$f(red)$} & 
\multicolumn{1}{c}{$D$} 
\\[2mm] \tableline\\
ACS-1 & 1.221(0.008) & 1.419(0.019) & 0.081(0.004) & 0.130(0.008) & 0.472(0.057) & 1.82(0.21) \\
ACS-3 & 1.262(0.016) & 1.503(0.029) & 0.095(0.007) & 0.129(0.011) & 0.456(0.086) & 2.12(0.20) \\
ACS-4 & 1.237(0.014) & 1.481(0.033) & 0.089(0.007) & 0.136(0.013) & 0.443(0.087) & 2.13(0.29) \\
ACS-5 & 1.204(0.010) & 1.409(0.023) & 0.087(0.005) & 0.149(0.006) & 0.381(0.058) & 1.69(0.15) \\
ACS-6 & 1.215(0.016) & 1.397(0.046) & 0.089(0.010) & 0.145(0.015) & 0.357(0.110) & 1.51(0.40) \\
WPC2  & 1.233(0.022) & 1.649(0.075) & 0.174(0.014) & 0.094(0.025) & 0.091(0.069) & 2.97(0.53) \\
\\[2mm] \tableline
\end{tabular}
\end{center}
\vspace{0.4cm}
\end{table*}

The distribution of GC colors (CDF) is driven to first order by the underlying
\emph{metallicity distribution function} (MDF).  As noted above, in most galaxies
the CDF and MDF are bimodal, with a canonical metal-poor (MP) ``blue'' sequence centered 
near $\langle$[Fe/H]$\rangle \simeq -1.5$
and a metal-rich (MR) ``red'' sequence near $\langle$[Fe/H]$\rangle \simeq -0.5$
\citep[e.g.][among many others]{zepf_ashman93,forbes_etal97,forbes_etal2011,brodie_strader06,arnold_etal2011,blom_etal2012a,cantiello_etal2014,brodie_etal2014,kartha_etal2016}.
A numerically convenient way to characterize the two modes is by their Gaussian mean colors
($\mu_1, \mu_2$) (blue, red) and dispersions ($\sigma_1, \sigma_2$), as well as the fraction of
total number of GCs in each mode, ($f_1, f_2$) where $f_1 + f_2 \equiv 1$.  

For each field in this study, a bimodal-Gaussian
fit to the CDF was done with the GMM code \citep{muratov_gnedin2010}.  The color histograms
are displayed in Figure \ref{fig:gmm}, and 
in Table \ref{tab:gmm} the details for the GMM fits are summarized. 
The last column gives the statistic $D = (\mu_2-\mu_1)/\sqrt{(\sigma_1^2+\sigma_2^2/2}$, which is an estimate 
of the degree of separation of the two modes 
(bimodality favored if $D \gtrsim 2$).  The data included for these fits are the same 
datasets restricted by magnitude and color as defined in Section 3.4 above.

The five ACS fields give highly consistent results for the mean colors
$\mu_1, \mu_2$ and dispersions $\sigma_1, \sigma_2$ of the blue and red modes.
The $D$ statistic is near $D \sim 2$ and the raw 
histograms are asymmetric for all the fields, suggesting that
intrinsic bimodality is preferred over unimodality.
In addition, the $p-$statistic giving the probability for unimodality
	(the null hypothesis) is $p \ll 0.01$ for all 5 ACS fields, strongly
favoring bimodality.
The red (metal-richer) fraction $f_2 = f_{red}$ decreases
gradually outward from the Perseus center, as expected by comparison with
Virgo and Coma \citep{durrell_etal2014,peng_etal2011}.  That is, metal-richer (redder) GCs are
more centrally concentrated within galaxies and are in any case not as numerous
as the blue (metal-poor) ones in the smaller galaxies that should have been
the main progenitors for the ICL.

Converting the $(F555W-F814W)$ values in Table \ref{tab:gmm} to intrinsic
color $(V-I)_0$ for the 5 ACS fields gives mean dereddened values
$(V-I)_0 = 0.92$ (blue mode) and 1.11 (red mode).  These are both within the ranges
of $(\mu_1, \mu_2$) for normal early-type galaxies; for example, \citet{larsen_etal2001}
find mean colors $(V-I)_0 = 0.90 - 0.99$ for the blue mode and $1.10 - 1.21$ for the
red mode, systematically increasing with host galaxy luminosity.  Again, if the ICL is dominated
by material stripped from $L_{\star}-$type galaxies (see above), then we would expect
the mean colors $\mu_1, \mu_2$ both to be a bit bluer than for the giant galaxies.

The results for the WFPC2 field are considerably more uncertain because of the
shallower magnitude limit and smaller sample size. The blue component
lies at $\mu_1 = 1.23$, corresponding to $(V-I)_0 = 1.04$, which is between
the standard blue and red modes described above.  The nominal red component $\mu_2$ is
small (only 9\% of the total) and we cannot rule out that it is due simply to 
some residual field contamination (see above) combined with greater photometric error than in the ACS fields.  

For the combined WFPC2 fields, the GMM solution gives $p = 0.055$, which does not
strongly favor intrinsic bimodality.  In addition, the blue-mode
dispersion $\sigma_1 = 0.17$ is twice as large as the blue-mode dispersions in the 
deeper ACS fields.  The main population $f_1$ may therefore simply be a combination of
the normal blue and red modes blurred together by the higher photometric uncertainty.
If we restrict the sample further to $F814W < 23.5$ as in Section 3.4 above, a GMM
fit yields similar results:  the nominal blue and red modes are not separated
and the color distribution is not clearly different from unimodal. 
Deeper and more precise photometry will be required to reach a
clearer answer. 

\section{NGC 1272 and 1275}

\begin{figure}[t]
\vspace{0.5cm}
\begin{center}
\includegraphics[width=0.5\textwidth]{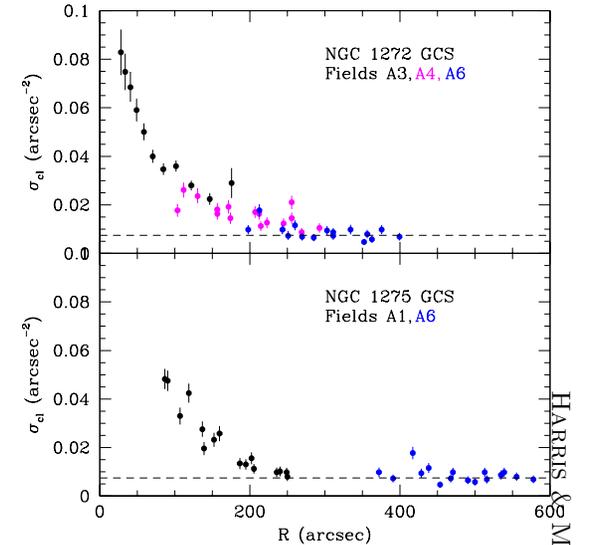}
\end{center}
\vspace{-0.0cm}
\caption{\emph{Upper panel:} Radial distribution of unresolved (starlike) objects centered on NGC 1272,
	within the magnitude range $21 < F814W < 25.5$ and color range $1.0 < (F555W-F814W) < 1.8$
	(see text). Data from fields ACS-3 (black symbols), ACS-4 (magenta), and ACS-6 (blue) are shown here.
	The projected density per unit area $\sigma_{cl}$ is plotted versus projected
	radius $R$ from the center of NGC 1272. The \emph{dashed line} at $\sigma = 0.0075$ is the adopted
	background level.
	\emph{Lower panel:} Radial distribution for unresolved objects in 
fields ACS-1 (black) and ACS-6 (blue), where $R$ is now the radius from the center of NGC 1275.}
\vspace{0.0cm}
\label{fig:n1272}
\end{figure}

\begin{figure}[t]
\vspace{-0.5cm}
\begin{center}
\includegraphics[width=0.5\textwidth]{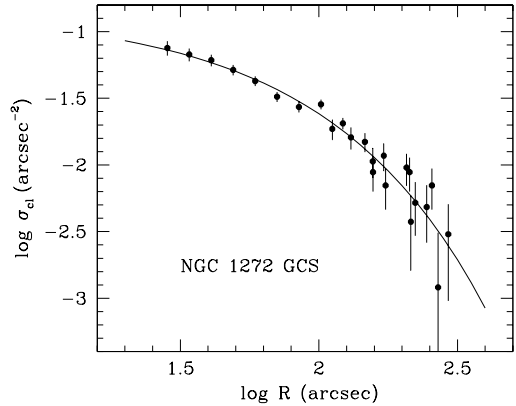}
\end{center}
\vspace{-0.0cm}
\caption{Radial distribution of the GC system around NGC 1272, after background subtraction.
	The datapoints include the numbers from ACS-3 and 4.
The best-fit S\'ersic function as described in the text is shown as the solid line.}
\vspace{0.0cm}
\label{fig:sersic}
\end{figure}

The placement of the ACS fields allows a good opportunity to determine the spatial distribution
of the NGC 1272 GC system in particular.  ACS-3 contains the galaxy, while ACS-4 and ACS-6 
are immediately adjacent and (helpfully) sit on the opposite side from NGC 1275.
(ACS-5 is equidistant from both NGC 1272 and 1275 and 
its GC population might be expected to comprise a roughly equal mixture
of the two; see again Fig.~\ref{fig:radial}.)  In Figure \ref{fig:n1272} (upper panel), the number density
$\sigma_{cl}$ (objects per arcsec$^{-2}$) is plotted for ACS-3,4,6 relative to the center of 
NGC 1272.  Here $\sigma_{cl}$ decreases smoothly outward, converging to a near-constant
level at the largest radii sampled.  From the outermost 6 bins (containing 124 objects) in ACS-6 we 
adopt a background $\sigma_b = 0.0075 \pm 0.00067$ arcsec$^{-2}$, indicated by the dashed line
in Fig.~\ref{fig:n1272}.  As noted above, this $\sigma_b$ is only slightly larger than the
level of the inner 
the WFPC2 fields (Fig.~\ref{fig:radial}), suggesting again that ACS-6 is sampling primarily
the IGC population.  Subtracting this $\sigma_b$ from the radial profile gives the result shown
in Figure \ref{fig:sersic}.
A S\'ersic-type function \citep{sersic1968} given by
\begin{equation}
	        \sigma_{cl} \, = \, \sigma_e {\rm exp}(-b_n [({R \over R_e})^{1/n} - 1] )
	\end{equation}
matches the result accurately, with best-fit parameters $n = 1.3$, $R_e = 153''$,
$\sigma_e = 0.0122$ arcsec$^{-2}$, $b_n = 2.46$.

Integration of the model profile out to $R \simeq 500''$ yields a total of $N=3570 \pm 150$ GCs
brighter than our adopted limit $F814W = 25.5$.  This limit is 0.7 mag brighter than the
expected GCLF turnover point $F814W(to) = 26.2$ (Fig.~\ref{fig:lf}).
Extrapolating in turn to find the total number of 
GCs over all magnitudes then yields $N_{GC} = 12000 \pm 1400$, where the stated uncertainty 
includes the Poisson count statistics, a $\pm0.1-$mag uncertainty in the GCLF turnover
luminosity, and a $\pm 0.1-$mag uncertainty in the GCLF dispersion.
Finally, the GCS specific frequency \citep{harris_vandenbergh1981}
is $S_N = N_{GC} \cdot 10^{0.4(M_V^T+15)} = 7.9 \pm 0.9$, a moderately
high value within the typical range for supergiant ellipticals \citep[cf.][]{harris_etal2013}.
A remaining question is whether or not the GCS from NGC 1275 itself is contributing to
the NGC 1272 totals particularly in ACS-3.  From the NGC 1275 profile (see comments below and
the lower panel of Fig.~\ref{fig:n1272}), it appears that the contribution from
NGC 1275 \emph{at the $300'' =$ 110 kpc projected distance of NGC 1272} will be at
the level of only a few percent.  If and when more complete imaging coverage of the core
Perseus region is available, it will be possible to solve for the spatial distributions of
both galaxies simultaneously and more accurately \citep[see e.g.][for an example in the Hydra cluster]{wehner_etal2008}.

For the central giant NGC1275, the available material is much more limited
and the conclusions about total GC population correspondingly rougher.
We use field ACS-1 along with ACS-6 to set the local background, though here we are forced to assume
that $\sigma_b$ is roughly constant over the region concerned.  The measurements are shown
in the lower panel of Fig.~\ref{fig:n1272}, replotted from Fig.~\ref{fig:radial}. 
The combination of this radial profile and
the inner data from \citet{penny_etal2012} indicates that $\sigma_{cl} \sim R^{-1.6}$ from $R \simeq 60''$
outward, and is relatively flat inside $60''$.  
The main concerns in estimating its total population are the true local background level,
and how far outward the NGC 1275 halo extends before fading into the ICM; the present data do
not answer either of these satisfactorily.  Here we simply use  
an outer radius $R_{max} = 270''$ (100 kpc) for integration of the profile 
and $\sigma_b \simeq 0.0075 \pm 0.00067$ arcsec$^{-2}$ as above.
The result is then $N = 3990 \pm 380$ GCs for $F814W < 25.5$,
not including any (unknown) systematic error in the adopted background.
Extrapolating over all magnitudes as described above gives
$N_{GC} = 13500 \pm 2000$ and a specific frequency $S_N = 12.2 \pm 1.8$,
including $\pm0.1-$mag uncertainties in the GCLF turnover and dispersion
as above.
This total GC population and $S_N$ would make NGC 1275 quite similar to M87 in Virgo
\citep[e.g.][]{harris2009b}.

We emphasize once again that a better assessment will
require a wider-field survey in which the GC populations in the member 
galaxies, and the IGCs, can both be determined more completely.
Nevertheless, this very limited imaging material is enough to 
suggest that 
both of the dominant Perseus galaxies appear to have very populous GC systems.
The closest analogy may be the more well known Coma cluster, where both NGC 4874
and 4889 have similarly large GC populations but where the central cD (NGC 4874)
has the higher specific frequency.

\section{Discussion: The IGC Population}

The data discussed in this paper consist of only 11 pencil-beam probes of the
GC population, and thus fall far short of a genuine wide-field survey
of the Perseus cluster.  Only a more comprehensive survey to similar depth will
be able to address the true spatial distribution of the IGCs and their total number.

Nevertheless, we can make a \emph{very rough} attempt to gauge the total GC population.
We assume, following Figs.~\ref{fig:radial} and \ref{fig:n1272},  
that $\sigma(IGC)$ is approximately constant at $\sigma_0 \simeq 0.0078$ arcsec$^{-2}$
in the inner part of the cluster and then falls off as $\sim R^{-1.2}$ further
out.  The inflection point where the power law reaches $\sigma_0$ is at
$R = 500''$ = 180 kpc from cluster center.  Integrating $\sigma(IGC)$ out to
the most remote field, WFPC2-6 at $R = 0.38^o$ = 500 kpc, then gives
$N \simeq 24,000$ GCs brighter than $F814W=25.5$.  
Extrapolating over the entire GCLF 
yields $N_{IGC} \sim 82000$ IGCs over this radial range.            
To emphasize its limitations once again, it is obvious that this argument
works forward from an extremely limited sampling of the Perseus region, so it cannot
account for any details and asymmetries of the IGM spatial structure.

In Coma \citet{peng_etal2011} trace the IGC component out to $R \sim 500$ kpc and find
that it must fall rapidly outside of that.  \citet{durrell_etal2014} measure the
Virgo IGCs out to $R \simeq 700$ kpc from the center of M87, again finding that their
number density falls rapidly after that.  The Perseus results, though quite
tentative, fall into the same general pattern. 

A better way to estimate the total GC population in Perseus at this stage may be to
use the mass ratio $\eta_M \equiv (M_{GCS}/{M_h})$ where $M_{GCS}$ is the total mass in
all the GCs and $M_h$ is the total mass in the entire cluster, dark plus baryonic (roughly,
its virial mass $M_{200}$).  Empirically, $\eta_M$ has been found to be nearly constant
over a wide range of environments including dwarf or giant galaxies,
spirals or ellipticals, and entire galaxy clusters; see, e.g., 
\citet{blakeslee_etal1997,spitler_forbes2009,durrell_etal2014,hudson_etal2014,harris_etal2015,harris_etal2017}.
The most recent calibration gives $\langle \eta_M \rangle = (4.3 \pm 0.6) \times 10^{-5}$
for entire clusters of galaxies 
\citep{harris_etal2017}.  For the Perseus cluster, $M_{200} = (6.65 \pm 0.45) \times 10^{14} M_{\odot}$
from \citet{simionescu_etal2011} determined from its X-ray gas pressure profile.  
If we then take a mean GC mass $M_{GC} = 2.8 \times 10^5 M_{\odot}$ appropriate for
$L_{\star}$ galaxies \citep{harris_etal2017}, 
the resulting estimate of the total population is $N_{GC} = 102,000$ clusters. 
If analogies with Virgo or Coma are useful, roughly half of these should belong to the
member galaxies, with the remainder in the IGM; thus, $N_{IGC} \sim 50,000$,
which agrees to within a factor of two with our estimate working directly from
the observations.
GC populations this large are not unprecedented:
Coma, which has a total mass very similar to Perseus, contains 
at least $N_{GC} = 70,000$ GCs \citep{peng_etal2011}, and estimates
well above $N_{GC} = 100,000$ have been proposed for 
the more distant clusters A1689 and A2744 \citep{alamo-martinez_etal2013,lee_jang2016}.

This range of estimates for $N_{GC}$, though admittedly crude, indicates that the Perseus IGC component
may be quite comparable with the Virgo and Coma clusters, and reinforces the growing
evidence that rich clusters of galaxies contain large amounts of intragalactic stellar light.
Though far more dilute, the ICL may add up to as much stellar mass as is in the galaxies, and
is the visible tracer of a long history of galaxy interaction and tidal stripping.

\section{Summary}

We have used photometry from HST ACS and WFPC2 imaging within the rich Perseus cluster (Abell 426) 
to isolate its
globular cluster population and to attempt detection of an Intragalactic population.
Our principal findings are these:
\begin{enumerate}
\item{} In the target fields, GC populations are clearly present following a
	normal Gaussian distribution versus magnitude and a bimodal distribution in 
	$(V-I)$ color that matches the expected properties seen in individual galaxies.
\item{} In the core region of Perseus, large numbers of GCs are present, but these
	are dominated by the halos of the two largest galaxies in the region, NGC 1275 and NGC 1272.
	We estimate that NGC 1272 contains $\simeq 12000$ GCs with $S_N = 7.9 \pm 0.9$,
	while for the central giant NGC 1275 we estimate more crudely $N_{GC} \sim 13500$
	and $S_N \sim 12$, very similar to M87.
\item{} The radial distribution of GCs counts indicates that an IGC component is present
	extending several hundred kpc away from the Perseus center.  Two different methods
	lead to the (very tentative) conclusion that there may be $\sim 50000 - 80000$ IGCs in 
Perseus as a whole, which would make it comparable with the Virgo and Coma clusters.  
\item{} A more comprehensive, wider-field survey of the Perseus cluster promises
	to be rewarded with a much more complete understanding of the Intragalactic cluster 
	population and the entire ICL component.
\end{enumerate}

\acknowledgements

We thank an anonymous referee for useful comments.
WEH acknowledges financial support 
from NSERC (Natural Sciences and Engineering Research Council of Canada).

\vspace{5mm}
\facilities{HST (ACS, WFPC2)}


\label{lastpage}

\end{document}